\documentclass[letterpaper,preprint,english,showpacs,pra]{revtex4}
\usepackage[T1]{fontenc}
\usepackage[latin1]{inputenc}
\usepackage{graphicx}
\usepackage{amssymb}

\makeatletter

\newcommand{\job}{J. Opt. B: Quant. Semiclass. Opt. }
\newcommand{\pr}{Phys. Rev. }
\newcommand{\pla}{Phys. Lett. A }
\newcommand{\UQ}{ARC Centre of Excellence for Quantum-Atom Optics, School of Physical Sciences, University of Queensland, Brisbane, Qld 4072, Australia.} 
 
\usepackage{graphicx}
\usepackage{bm}
\usepackage{amsmath}
\usepackage{babel}
\makeatother
\begin{document}

\title{Entanglement and the Einstein-Podolsky-Rosen paradox with coupled
intracavity optical downconverters}

\author{M.~K. Olsen and P.~D. Drummond}

\affiliation{\UQ{}}

\date{\today{}}

\begin{abstract}
We show that two evanescently coupled $\chi^{(2)}$ parametric downconverters
inside a Fabry-Perot cavity provide a tunable source of quadrature
squeezed light, Einstein-Podolsky-Rosen correlations and quantum entanglement.
Analysing the operation in the below threshold regime, we show how
these properties can be controlled by adjusting the coupling strengths
and the cavity detunings. As this can be implemented with integrated
optics, it provides a possible route to rugged and stable EPR sources. 
\end{abstract}


\maketitle

\section{Introduction}

The Einstein, Podolsky and Rosen (EPR) paradox stems from a famous
paper published in $1935$~\cite{EPR}, which showed that local realism
is not consistent with quantum mechanical completeness. A direct and
feasible demonstration of the EPR paradox with continuous variables
was first suggested using nondegenerate parametric amplification (also
known as the OPA)~\cite{eprquad}. The optical quadrature phase amplitudes
used in these proposals have the same mathematical properties as the position and momentum
originally used by EPR. Even though the correlations between these
are not perfect, they are still entangled sufficiently to allow for
an inferred violation of the uncertainty principle, which is equivalent
to the EPR paradox~\cite{eprMDR,rd}. An experimental demonstration
of this proposal by Ou \emph{et al.} soon followed, showing a clear
agreement with quantum theory~\cite{Ou}.

In this work, rather than using the nondegenerate optical parametric oscillator (OPO),  we consider an alternative device using two 
degenerate type I downconverters 
inside the same optical cavity,
and coupled by evanescent overlaps of the intracavity modes within the nonlinear medium. Generally, such a device may be considered as either a 
single nonlinear crystal pumped 
by two spatially separated lasers, or two waveguides with a $\chi^{(2)}$ component. We
calculate phase-dependent correlations between the two low frequency
outputs of the cavity in the below threshold regime, showing that
this system exhibits a wide range of behaviour and is potentially
an easily tunable source of single-mode squeezing, entangled states
and states which exhibit the EPR paradox. The spatial separation of
the output modes means that they do not have to be separated by optical
devices before measurements can be made, along with the unavoidable
losses which would result from this procedure. The entangled beams
produced can be degenerate in both frequency and polarisation, unlike
those of the nondegenerate OPO, and would exit the cavity at spatially
separated locations. These correlations are tunable by controlling some of the
operational degrees of freedom of the system, including the evanescent
couplings between the two waveguides, the input power and the cavity
detunings. 

The term nonlinear coupler was given to a system of two coupled waveguides
without an optical cavity by Pe\u{r}ina \emph{et al.\/{}}~\cite{coupler}.
Generically, the device consists of two parallel optical waveguides
which are coupled by an evanescent overlap of the guided modes. The
quantum statistical properties of this device when the nonlinearity
is of the $\chi^{(3)}$ type have been theoretically investigated,
predicting energy transfer between the waveguides~\cite{Ibrahim}
and the generation of correlated squeezing~\cite{korea}. Coupled
$\chi^{(2)}$ downconversion processes in the travelling wave configuration
have also been examined theoretically, predicting that light
produced in one of the media can be controlled by light entering the
other~\cite{mista}, and that such a device can produce entanglement
of the output beams~\cite{herec}. The coupler with $\chi^{(2)}$
nonlinearity held inside a pumped Fabry-Perot cavity, and operating
in the second harmonic generation (SHG) configuration, was introduced
by Bache \emph{et al.\/{}}~\cite{dimer}, who named it the quantum
optical dimer by analogy with various systems that display coupling
between discrete sites. They analysed intensity correlations between
the modes, predicting noise suppression in both the sum and the difference.

As the intracavity $\chi^{(2)}$ downconversion processes have long
been appreciated as sources of quantum states of the electromagnetic
field (See Martinelli \emph{et al.\/{}}~\cite{bigpaul} for an overview),
we will combine and extend these previous analyses to consider two
coupled downconverters operating inside a Fabry-Perot cavity. 
The advantage of this proposal is the all-integrated nature of the
device, which promises greatly increased robustness. Additional potential
advantages are the reductions in threshold pump power and phase noise,
relative to current practise. Another potential advantage as compared to the normal type II
polarisation nondegenerate OPO lies in the difficulty of 
stabilising this device at frequency degeneracy~\cite{Fabre1,Fabre2}. Our proposal should be well stabilised
by the linear coupling, without having to add any additional features.

\section{The system and equations of motion}
\label{sec:equations}

The physical device we wish to examine differs from that described
in Ref.~\cite{dimer} in one important detail. We will analyse it
in the downconversion regime, where the cavity pumping is at a frequency
$2\omega_{L}\simeq\omega_{b}$. As this device has been described
in detail in Ref.~\cite{dimer}, we will give a briefer description
of the essential features here. The system of interest consists of two
coupled nonlinear $\chi^{(2)}$ waveguides inside a driven optical
cavity, which may utilise integrated Bragg reflection for compactness.
Each waveguide supports two resonant modes at frequencies $\omega_{a},\,\omega_{b}$,
where $2\omega_{a}\simeq\omega_{b}$. The higher frequency modes at
$\omega_{b}$ are driven coherently with an external laser, while
the nonlinear interaction within the waveguides produces pairs of
downconverted photons with frequency $\omega_{a}$ . We assume that
only the cavity modes at these two frequencies are important and that
there is perfect phase matching inside the media. The two waveguides
are evanescently coupled as in Ref.~\cite{dimer}. Besides the differences
in the pumping frequency, we will be interested in the phase-dependent
correlations necessary to demonstrate entanglement and the EPR paradox,
rather than the intensity correlations of Ref.~\cite{dimer}.

The effective Hamiltonian for the system can be written as 
\begin{equation}
{\mathcal{H}}_{eff}={\mathcal{H}}_{int}+{\mathcal{H}}_{couple}+{\mathcal{H}}_{pump}+{\mathcal{H}}_{res},
\label{eq:Heff}
\end{equation}
where the nonlinear interaction with the $\chi^{(2)}$ media is described
by 
\begin{equation}
{\mathcal{H}}_{int}=i\hbar\frac{\kappa}{2}\left[\hat{a}_{1}^{\dag\;2}\hat{b}_{1}-\hat{a}_{1}^{2}\hat{b}_{1}^{\dag}+
\hat{a}_{2}^{\dag\;2}\hat{b}_{2}-\hat{a}_{2}^{2}\hat{b}_{2}^{\dag}\right]\,\,.
\label{eq:Hnl}
\end{equation}
Here $\kappa$ denotes the effective nonlinearity of the waveguides
(we assume that the two are equal), and $\hat{a}_{k},\;\hat{b}_{k}$
are the bosonic annihilation operators for quanta at the frequencies
$\omega_{a},\;\omega_{b}$ within the crystal $k\;(=1,2)$. The coupling
by evanescent waves is described by 
\begin{equation}
{\mathcal{H}}_{couple}=\hbar J_{a}\left[\hat{a}_{1}\hat{a}_{2}^{\dag}+\hat{a}_{1}^{\dag}\hat{a}_{2}\right]+\hbar J_{b}
\left[\hat{b}_{1}\hat{b}_{2}^{\dag}+\hat{b}_{1}^{\dag}\hat{b}_{2}\right],
\label{eq:Hcouple}
\end{equation}
where the $J_{k}$ are the coupling parameters at the two frequencies,
as described in Ref.~\cite{dimer}. We note that in that work it
is stated that the lower frequency coupling, $J_{a}$, is generally
stronger than the higher frequency coupling, $J_{b}$, and also that
values of $J_{a}$ as high as $50$ times the lower frequency cavity
loss rate were calculated to be physically reasonable. The cavity
pumping is described by 
\begin{equation}
{\mathcal{H}}_{pump}=i\hbar\left[\epsilon_{1}\hat{b}_{1}^{\dag}-\epsilon_{1}^{\ast}\hat{b}_{1}+\epsilon_{2}\hat{b}_{2}^{\dag}-
\epsilon_{2}^{\ast}\hat{b}_{2}\right],
\label{eq:Hpump}
\end{equation}
where the $\epsilon_{k}$ represent pump fields which we will describe
classically. Finally, the cavity damping is described by 
\begin{equation}
{\mathcal{H}}_{res}=\hbar\sum_{k=1}^{2}\left(\Gamma_{a}^{k}\hat{a}_{k}^{\dag}+\Gamma_{b}^{k}\hat{b}_{k}^{\dag}\right)+h.c.,
\label{eq:Hres}
\end{equation}
where the $\Gamma^{k}$ represent bath operators at the two frequencies
and we have made the usual zero temperature approximation for the
reservoirs.

With the standard methods~\cite{GardinerQN}, and using the operator/c-number
correspondences $(\hat{a}_{j}\leftrightarrow\alpha_{j},\hat{b}_{j}\leftrightarrow\beta_{j})$,
the Hamiltonian can be mapped onto a Fokker-Planck equation for the
Glauber-Sudarshan P-distribution~\cite{Roy,Sud}. However, as the
diffusion matrix of this Fokker-Planck equation is not positive-definite,
it cannot be mapped onto a set of stochastic differential equations.
Hence we will use the positive-P representation~\cite{plusP} which,
by doubling the dimensionality of the phase-space, allows a Fokker-Planck
equation with a positive-definite diffusion matrix to be found and
thus a mapping onto stochastic differential equations. Making the
correspondence between the set of operators $(\hat{a}_{j},\hat{a}_{j}^{\dag},\hat{b}_{j},\hat{b}_{j}^{\dag})$
$(j=1,2)$ and the set of c-number variables $(\alpha_{j},\alpha_{j}^{+},\beta_{j},\beta_{j}^{+})$,
we find the following set of equations, 
\begin{eqnarray}
\frac{d\alpha_{1}}{dt} & = & -(\gamma_{a}+i\Delta_{a})\alpha_{1}+\kappa\alpha_{1}^{+}\beta_{1}+iJ_{a}\alpha_{2}+\sqrt{\kappa\beta_{1}}\;\eta_{1}(t),\nonumber \\
\frac{d\alpha_{1}^{+}}{dt} & = & -(\gamma_{a}-i\Delta_{a})\alpha_{1}^{+}+\kappa\alpha_{1}\beta_{1}^{+}-iJ_{a}\alpha_{2}^{+}+
\sqrt{\kappa\beta_{1}^{+}}\;\eta_{2}(t),\nonumber \\
\frac{d\alpha_{2}}{dt} & = & -(\gamma_{a}+i\Delta_{a})\alpha_{2}+\kappa\alpha_{2}^{+}\beta_{2}+iJ_{a}\alpha_{1}+\sqrt{\kappa\beta_{2}}\;\eta_{3}(t),\nonumber \\
\frac{d\alpha_{2}^{+}}{dt} & = & -(\gamma_{a}-i\Delta_{a})\alpha_{2}^{+}+\kappa\alpha_{2}\beta_{2}^{+}-iJ_{a}\alpha_{1}^{+}+
\sqrt{\kappa\beta_{2}^{+}}\;\eta_{4}(t),\nonumber \\
\frac{d\beta_{1}}{dt} & = & \epsilon_{1}-(\gamma_{b}+i\Delta_{b})\beta_{1}-\frac{\kappa}{2}\alpha_{1}^{2}+iJ_{b}\beta_{2},\nonumber \\
\frac{d\beta_{1}^{+}}{dt} & = & \epsilon_{1}^{\ast}-(\gamma_{b}-i\Delta_{b})\beta_{1}^{+}-\frac{\kappa}{2}\alpha_{1}^{+\;2}-iJ_{b}\beta_{2}^{+},\nonumber \\
\frac{d\beta_{2}}{dt} & = & \epsilon_{2}-(\gamma_{b}+i\Delta_{b})\beta_{2}-\frac{\kappa}{2}\alpha_{2}^{2}+iJ_{b}\beta_{1},\nonumber \\
\frac{d\beta_{2}^{+}}{dt} & = & \epsilon_{2}^{\ast}-(\gamma_{b}-i\Delta_{b})\beta_{2}^{+}-\frac{\kappa}{2}\alpha_{2}^{+\;2}-iJ_{b}\beta_{1}^{+},
\label{eq:PPSDE}
\end{eqnarray}
where the $\gamma_{k}$ represent cavity damping. We have also added
cavity detunings $\Delta_{a,b}$ from the two resonances, so that
for a pump laser at angular frequency $2\omega_{L}$, one has $\Delta_{a}=\omega_{a}-\omega_{L}$and
$\Delta_{b}=\omega_{b}-2\omega_{L}$. Below, in section~\ref{sec:detune},
we will investigate detuning effects in greater detail. The real Gaussian
noise terms have the correlations $\overline{\eta_{j}(t)}=0$ and
$\overline{\eta_{j}(t)\eta_{k}(t')}=\delta_{jk}\delta(t-t')$. Note
that, due to the independence of the noise sources, $\alpha_{k}\;(\beta_{k})$
and $\alpha_{k}^{+}\;(\beta_{k}^{+})$ are not complex conjugate pairs,
except in the mean over a large number of stochastic integrations
of the above equations. However, these equations do allow us to calculate
the expectation values of any desired time-normally ordered operator
moments, exactly as required to calculate spectral correlations.

\section{Linearised analysis}
\label{sec:linearise}

In an operating regions where it is valid, a linearised fluctuation
analysis provides a simple way of calculating both intracavity and
output spectra of the system~\cite{DFW,mjc}, by treating it as an
Ornstein-Uhlenbeck process~\cite{ornstein}. To perform this analysis
we first divide the variables of Eq.~\ref{eq:PPSDE} into a steady-state
mean value and a fluctuation part, e.g. $\alpha_{1}\rightarrow\alpha_{1}^{ss}+\delta\alpha_{1}$
and so on for the other variables. We find the steady state solutions
by solving the equations (\ref{eq:PPSDE}) without the noise terms (note that in this section we will treat all fields as being at resonance),
and write the equations for the fluctuation vector 
$\delta\tilde{x}=[\delta\alpha_{1},\delta\alpha_{1}^{+},\delta\alpha_{2},\delta\alpha_{2}^{+},\delta\beta_{1},\delta\beta_{1}^{+},\delta\beta_{2},\delta\beta_{2}^{+}]^{T}$,
to first order in these fluctuations, as 
\begin{equation}
d\;\delta\tilde{x}=-A\delta\tilde{x}\; dt+BdW,
\label{eq:dAB}
\end{equation}
where the drift matrix is 
\begin{equation}
A=\left[\begin{array}{cc}
A_{aa} & -A_{ba}^{*}\\
A_{ba} & A_{bb}\end{array}\right],
\end{equation}
with
\begin{eqnarray}
A_{aa} & = & \left[\begin{array}{cccc}
\gamma_{a} & -\kappa\beta_{1}^{ss} & -iJ_{a} & 0\\
-\kappa\beta_{1}^{ss\ast} & \gamma_{a} & 0 & iJ_{a}\\
-iJ_{a} & 0 & \gamma_{a} & -\kappa\beta_{2}^{ss}\\
0 & iJ_{a} & -\kappa\beta_{2}^{ss\ast} & \gamma_{a}\end{array}\right],\nonumber \\
A_{ba} & = & \left[\begin{array}{cccc}
\kappa\alpha_{1}^{ss} & 0 & 0 & 0\\
0 & \kappa\alpha_{1}^{ss\ast} & 0 & 0\\
0 & 0 & \kappa\alpha_{2}^{ss} & 0\\
0 & 0 & 0 & \kappa\alpha_{2}^{ss\ast}\end{array}\right],\nonumber \\
A_{bb} & = & \left[\begin{array}{cccc}
\gamma_{b} & 0 & -iJ_{b} & 0\\
0 & \gamma_{b} & 0 & iJ_{b}\\
-iJ_{b} & 0 & \gamma_{b} & 0\\
0 & iJ_{b} & 0 & \gamma_{b}\end{array}\right].
\end{eqnarray}
In this equation, $dW$ is a vector of real Wiener increments, and
the matrix $B$ is zero except for the first four diagonal elements,
which are respectively $\sqrt{\kappa\beta_{1}^{ss}},\;\sqrt{\kappa\beta_{1}^{ss\ast}},\;\sqrt{\kappa\beta_{2}^{ss}},\;\sqrt{\kappa\beta_{2}^{ss\ast}}$.
The essential conditions for this expansion to be valid are that moments
of the fluctuations be smaller than the equivalent moments of the
mean values, and that the fluctuations stay small. In the case of
the single optical parametric oscillator (OPO), it is well known that
there is a critical operating point around which this condition does
not hold. This point is easily found by examination of the eigenvalues
of the equivalent fluctuation drift matrix for that system, and this
procedure is also valid in the present case. The fluctuations will
not tend to grow as long as none of the eigenvalues of the matrix
$A$ develop a negative real part. At the point at which this happens
the linearised fluctuation analysis is no longer valid, as the fluctuations
can then grow exponentially and the necessary conditions for linearisation
are no longer fulfilled. In this work we will only be interested in
a region where linearisation is valid.

To examine the stability of the system, we first need to find the
steady state solutions for the amplitudes, by solving for the steady
state of Eq.~\ref{eq:PPSDE} with the noise terms dropped. As in
the usual optical parametric oscillator, there is an oscillation threshold
below which $\alpha_{j}^{ss}=0$ and only the high frequency mode
is populated. In the present case, for a real pump, we find $\beta_{j}^{ss}=\epsilon/(\gamma_{b}-iJ_{b})$.
Inserting these solutions in the matrix $A$ allows us to find simple
expressions for the eigenvalues, 
\begin{eqnarray}
\lambda_{1,2} & = & \gamma_{b}+iJ_{b},\nonumber \\
\lambda_{3,4} & = & \gamma_{b}-iJ_{b},\nonumber \\
\lambda_{5,6} & = & \gamma_{a}+\sqrt{\left[\kappa^{2}\epsilon^{2}/\widetilde{\gamma}_{b}^{2}-J_{a}^{2}\right]},\nonumber \\
\lambda_{7,8} & = & \gamma_{a}-\sqrt{\left[\kappa^{2}\epsilon^{2}/\widetilde{\gamma}_{b}^{2}-J_{a}^{2}\right]}.
\label{eq:autovalores}
\end{eqnarray}
Here we have introduced auxiliary variables, $\widetilde{\gamma}{}_{a,b}=\sqrt{\gamma_{a,b}^{2}+J_{a,b}^{2}}$.
We immediately see that $\lambda_{7,8}$ can develop negative real
parts for a pump amplitude greater than the critical value, $\epsilon_{c}=\widetilde{\gamma}{}_{a}\widetilde{\gamma}{}_{b}/\kappa$.
As it must, this expression reduces to the single OPO expression of
$\gamma_{a}\gamma_{b}/\kappa$ when the couplings are set to zero.
In that case, there is then a stable above threshold solution in which
the high frequency mode inside the cavity remains constant, independently
of any further increase in the pumping, and the low frequency mode
becomes occupied. 

In the present case, it is not simple to find general expressions
for these above threshold solutions analytically, but as we will concentrate
our attention on the rich variety of below threshold behaviour which
is exhibited, this is not important here. We note here that, unlike
the single OPO case with a resonant cavity, the threshold pumping
is not a constant for fixed cavity loss rates, but is a function of
the coupling strengths between the two waveguides. Using the below
threshold solutions, we may calculate any desired time normally-ordered
spectral correlations inside the cavity using the simple formula 
\begin{equation}
S(\omega)=\left(A+i\omega\openone\right)^{-1}BB^{\text{T}}\left(A^{\text{T}}-i\omega\openone\right)^{-1},
\label{eq:inspek}
\end{equation}
after which we use the standard input-output relations~\cite{mjc}
to relate these to quantities which may be measured outside the cavity.

\section{Quantum correlations}
\label{sec:correlations}

\subsection{Single mode squeezing}
\label{subsec:squeezado}

The first quantities we wish to calculate are the single mode quadrature
squeezing spectra, to compare these with the well-known results for
the normal uncoupled OPO. Defining the quadrature amplitudes as \begin{equation}
\hat{X}_{j}^{\theta}=\hat{a}_{j}\mbox{e}^{-i\theta}+\hat{a}_{j}^{\dag}\mbox{e}^{i\theta},\label{eq:Xthetageral}\end{equation}
 (where $j=1,2$), we will use the notation \begin{eqnarray}
\hat{X}_{j}^{0} & = & \hat{X}_{j},\nonumber \\
\hat{X}_{j}^{\frac{\pi}{2}} & = & \hat{Y}_{j}.\label{eq:XYgeral}\end{eqnarray}
 We note here that the quadrature definitions do not need to specify
whether it is mode $a$ or $b$ which is involved, as we do not find
any interesting behaviour in the high frequency modes below threshold
and hence will only present results for the low frequency modes. With
this normalisation the coherent state value for the quadrature variances
is one. To simplify our results we will assume that the pumping terms
for each crystal are real and equal $(\epsilon_{1}=\epsilon_{2}=\epsilon)$.

The expressions for the below threshold low frequency quadrature variances
in the single OPO case are well known~\cite{osenhor}, being 
\begin{eqnarray}
S_{X}^{\text{out}}(\omega) & = & 1+\frac{4\gamma_{a}\gamma_{b}\kappa\epsilon}{(\gamma_{a}\gamma_{b}-\kappa\epsilon)^{2}+\gamma_{b}^{2}\omega^{2}},\nonumber \\
S_{Y}^{\text{out}}(\omega) & = & 1-\frac{4\gamma_{a}\gamma_{b}\kappa\epsilon}{(\gamma_{a}\gamma_{b}+\kappa\epsilon)^{2}+\gamma_{b}^{2}\omega^{2}},
\label{eq:noncoupledVXVY}
\end{eqnarray}
and predicting zero-frequency squeezing which becomes perfect in
the $Y$ quadrature as the pump approaches the critical threshold
value, $\epsilon=\gamma_{a}\gamma_{b}/\kappa$, although the linearised analysis
breaks down near this point. Note that the variances inside and outside
the cavity are related by $S_{X}^{\text{out}}=1+2\gamma_{a}V(\hat{X})$.
Our coupled system would be expected to exhibit the above values in
the limit as $J_{a,b}\rightarrow 0$, which provides a standard for
comparison with the analytical results. In the general case, we find
that $S_{X_{1}^{\theta}}^{\text{out}}=S_{X_{2}^{\theta}}^{\text{out}}$,
as expected. We also find that the coupling means that the intracavity
high frequency field is no longer real, but has a phase given by $\Theta_{b}=\tan^{-1}(J_{b}/\gamma_{b})$. 

This will mean that the optimum correlations will no longer generally
be found in the $X_{j}$ and $Y_{j}$ quadratures, but at some other
phase angle, as found previously for second harmonic generation in
detuned cavities~\cite{granja}. Experimentally, this does not present
a problem as the local oscillator phase is normally swept across all
angles, which must therefore include the optimum angle. We can find
analytical solutions for the angle of maximum single-mode squeezing
(and antisqueezing), for example, these differing by $\pi/2$ and
being found as 
\begin{equation}
\theta_{opt}=\tan^{-1}\left\{ \frac{2V(\hat{X},\hat{Y})}{V(\hat{Y})-V(\hat{X})\pm\sqrt{\left[V(\hat{Y})-V(\hat{X})\right]^{2}+
4\left[V(\hat{X},\hat{Y})\right]^{2}}}\right\} ,
\label{eq:angulo}
\end{equation}
where $V(A,B)=\langle AB\rangle-\langle A\rangle\langle B\rangle$.
However, as this expression is a complicated function of several variables
when written out in full, and will not necessarily give the optimum
choices at all frequencies, nor when we consider correlations between
the modes, we will present results where the local oscillator angle
has been optimised numerically.

The $\hat{X}$ and $\hat{Y}$ spectral variances outside the cavity
are found as
\begin{eqnarray}
S_{X_{1,2}}^{\text{out}}(\omega) & = & 1+\frac{4\gamma_{a}\kappa\epsilon\left\{\gamma_{b}\left[\widetilde{\gamma_{b}}^{2}(\omega^{2}-J_{a}^{2})+
J_{b}^{2}\gamma_{a}^{2}+(\gamma_{a}\gamma_{b}+\kappa\epsilon)^{2}
\right]+2\gamma_{a}J_{b}^{2}\kappa\epsilon\right\}}
{4\gamma_{a}^{2}\widetilde{\gamma_{b}}^{4}\omega^{2}+\left[\widetilde{\gamma_{b}}^{2}(\widetilde{\gamma_{a}}^{2}-\omega^{2})-\kappa^{2}\epsilon^{2}\right]^{2}},\nonumber \\
S_{Y_{1,2}}^{\text{out}}(\omega) & = & 1-\frac{4\gamma_{a}\kappa\epsilon\left\{\gamma_{b}\left[\widetilde{\gamma_{b}}^{2}(\omega^{2}-J_{a}^{2})+
J_{b}^{2}\gamma_{a}^{2}+(\gamma_{a}\gamma_{b}-\kappa\epsilon)^{2}
\right]+2\gamma_{a}J_{b}^{2}\kappa\epsilon\right\}}
{4\gamma_{a}^{2}\widetilde{\gamma_{b}}^{4}\omega^{2}+\left[\widetilde{\gamma_{b}}^{2}(\widetilde{\gamma_{a}}^{2}-\omega^{2})-\kappa^{2}\epsilon^{2}\right]^{2}},
\label{eq:coupledVXVY}
\end{eqnarray}
which, as expected, reduce to the single OPO expressions above (\ref{eq:noncoupledVXVY})
when the coupling terms are set to zero. The output covariance is
\begin{equation}
V(\hat{X}_{j},\hat{Y}_{j})=\frac{4\gamma_{a}J_{b}\kappa\epsilon\left[\widetilde{\gamma}_{b}^{2}(\gamma_{a}^{2}-J_{a}^{2}+\omega^{2})+\kappa^{2}\epsilon^{2}\right]}
{4\gamma_{a}^{2}\widetilde{\gamma_{b}}^{4}\omega^{2}+\left[\widetilde{\gamma_{b}}^{2}(\widetilde{\gamma_{a}}^{2}-\omega^{2})-\kappa^{2}\epsilon^{2}\right]^{2}},
\label{eq:covXY}
\end{equation}
which will give $\theta_{opt}=0,\pi/2$ for the uncoupled case, where
$\hat{Y}$ is the squeezed quadrature and $\hat{X}$ the antisqueezed
quadrature.

\begin{figure}
\begin{center}\includegraphics[%
  width=0.90\columnwidth]{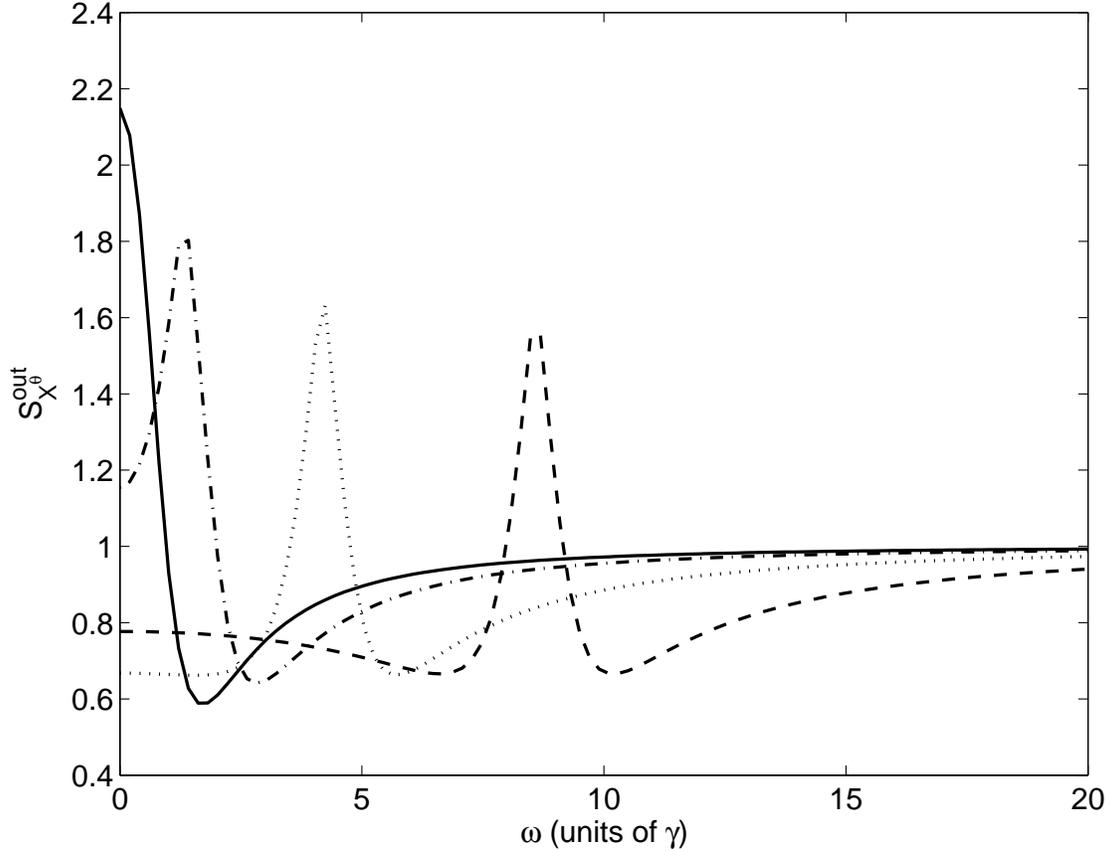}\end{center}

\caption{$S_{X^{\theta}}^{\text{out}}(\omega)$ for $\gamma=1$, $J_{b}=1$
and different $J_{a}$, all at the $\theta$ of maximum squeezing.
The solid line is for $J_{a}=1$ and $\theta=113^{o}$, the dash-dotted
line is $J_{a}=2$, the dotted line is $J_{a}=5$ and the dashed line
is $J_{a}=10$, all for $\theta=22^{o}$. The pump amplitude is $\epsilon=0.5\epsilon_{c}$
in each case and all quantities plotted in this and subsequent graphics
are dimensionless. Note that all plotted spectra are symmetric about
zero frequency and all results shown use the value $\kappa=0.01$ and $\gamma_{a}=\gamma_{b}=\gamma$.}

\label{fig:VXJbconst}
\end{figure}

In Fig.~\ref{fig:VXJbconst} we show the single-mode output spectral
quadrature variances for the quadrature of best squeezing as the low-frequency
mode coupling strength is varied, beginning with $J_{a}=J_{b}=\gamma_{a}=\gamma_{b}=\gamma$.
We note here that the pump values used in all the displayed results,
$\epsilon_{j}=0.5\epsilon_{c}$, depend on the couplings as stated
above and are therefore different for different combinations of the
couplings, but are all the same fraction of the threshold value. We
find less single-mode squeezing than in the uncoupled case for the
same ratio $\epsilon/\epsilon_{c}$, and also find that changing $J_{b}$
mainly serves to change the angle of maximum squeezing. Changing $J_{a}$
changes the frequency at which the maximum of squeezing is found.
We see that this device is not as efficient at producing squeezed single-mode
outputs as the normal OPO, but as we are interested in the quantum
correlations between the two output modes, we will now examine these.

\subsection{Entanglement and the EPR paradox}

\label{subsec:EPR}

An entanglement criterion for optical quadratures has been outlined
by Dechoum \emph{et al.\/{}}~\cite{ndturco}, following from criteria
developed by Duan \emph{et al.\/{}}~\cite{Duan} which are based
on the inseparability of the system density matrix. A theoretical
method to demonstrate the EPR paradox using quadrature amplitudes
was developed by Reid~\cite{eprMDR}, using the mathematical similarities
of the quadrature operators to the original position and momentum
operators. We will briefly outline these criteria here and then apply
them to our system, using the quadrature operators $\hat{X}_{j}$
and $\hat{Y}_{j}$. Note that even though these quadratures have the
same mathematical properties as the canonical position and momentum
operators for the harmonic oscillator, they correspond physically
to the real and imaginary parts of the electromagnetic field, not
its position and momentum.

To demonstrate entanglement between the modes, we define the combined
quadratures $\hat{X}_{\pm}=\hat{X}_{1}\pm\hat{X}_{2}$ and $\hat{Y}_{\pm}=\hat{Y}_{1}\pm\hat{Y}_{2}$
and calculate the variances in these, which we may do analytically.
Optimising the result for arbitrary phase angles is better performed
numerically. Following the treatment of Ref.~\cite{ndturco}, entanglement
is guaranteed provided that 
\begin{equation}
S_{X_{\pm}}^{\text{out}}+S_{Y_{\mp}}^{\text{out}}<4.
\label{eq:inequalityduan}
\end{equation}
We note here that the combined variance defined in this way has an
obvious relationship with the well-known two-mode squeezed states
which are produced, for example, by the nondegenerate OPO~\cite{democrat1,democrat2},
but that the quadratures between which we find entanglement here are
not the same as those which are entangled in that case, where these
are $\hat{X}_{-}$ and $\hat{Y}_{+}$. In the present case, considering
only the phase angles $\theta=0$ and $\pi/2$, we find entanglement
with $\hat{X}_{+}$ and $\hat{Y}_{-}$. The two individual variances
can be written as 
\begin{eqnarray}
S_{X_{\pm}}^{\text{out}} & = & S_{X_{1}}^{\text{out}}+S_{X_{2}}^{\text{out}}\pm2V(\hat{X}_{1},\hat{X}_{2}),\nonumber \\
S_{X_{\pm}}^{\text{out}} & = & S_{X_{1}}^{\text{out}}+S_{X_{2}}^{\text{out}}\pm2V(\hat{Y}_{1},\hat{Y}_{2}).
\label{eq:sumdiffvars}
\end{eqnarray}
The individual quadrature variances are given above (\ref{eq:coupledVXVY}),
while for the covariances we find: 
\begin{equation}
V(\hat{X}_{1},\hat{X}_{2})=\frac{-8J_{a}J_{b}\gamma_{a}^{2}\widetilde{\gamma}_{b}^{2}\kappa\epsilon}{{4\gamma_{a}^{2}\widetilde{\gamma}_{b}^{4}\omega^{2}
+\left[\widetilde{\gamma}_{b}^{2}(\widetilde{\gamma_{a}}^{2}-\omega^{2})-\kappa^{2}\epsilon^{2}\right]^{2}}},
\label{eq:VXX}
\end{equation}
and $V(\hat{Y}_{1},\hat{Y}_{2})=-V(\hat{X}_{1},\hat{X}_{2})$, showing
that the $\hat{X}$ quadratures are anticorrelated and the $\hat{Y}$
quadratures are correlated. Although these results allow us to write
analytical expressions for the combined variances, these are rather
bulky and not very enlightening, so we will not reproduce them here.

\begin{figure}
\begin{center}\includegraphics[%
  width=0.90\columnwidth]{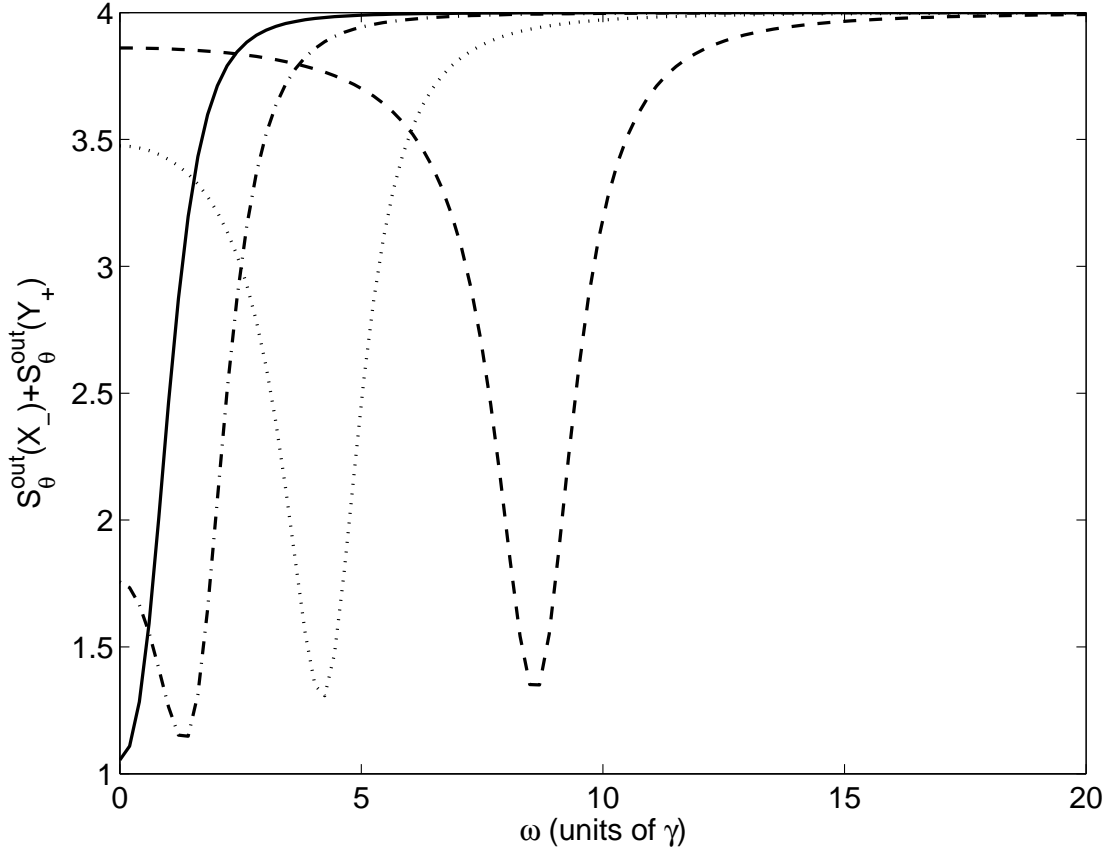}\end{center}

\caption{Demonstration of entanglement, using $S_{\theta}^{\text{out}}(\hat{X}_{-})+S_{\theta}^{\text{out}}(\hat{Y}_{+})$,
for $\gamma=1$, $J_{b}=1$, and $J_{a}=1$ (solid line), $2$ (dash-dotted
line), $5$ (dotted line) and $10$ (dashed line). The quadrature
angle for $\hat{X}$ is $67^{o}$ and that for $\hat{Y}$ is $157^{o}$.
The pump amplitude is $\epsilon=0.5\epsilon_{c}$.}

\label{fig:entangleJa}
\end{figure}

To optimise the degree of entanglement as a function of the quadrature
phase angle, we investigate the output spectral correlation 
\begin{equation}
S_{\theta}^{\text{out}}(\hat{X}_{-})+S_{\theta}^{\text{out}}(\hat{Y}_{+}),
\label{eq:Splusminus}
\end{equation}
where the $\hat{X}$ quadratures are at the angle $\theta$ and the
$\hat{Y}$ quadratures at the angle $\theta+\pi/2$. What we find,
as shown in Fig.~\ref{fig:entangleJa}, is that the degree of entanglement
and the frequency at which it exists depend on the coupling strength
$J_{a}$ while the optimum angle depends on $J_{b}$. When we hold
$J_{a}$ constant and increase $J_{b}$, we find that the maximum
of entanglement is always found at zero frequency, but that the optimum
quadrature angle changes.

To examine the utility of the system for the production of states
which exhibit the EPR paradox, we follow the approach of Reid~\cite{eprMDR}.
We assume that a measurement of the $\hat{X}_{1}$ quadrature, for
example, will allow us to infer, with some error, the value of the
$\hat{X}_{2}$ quadrature, and similarly for the $\hat{Y}_{j}$ quadratures.
This allows us to make linear estimates of the quadrature variances,
which are then minimised to give the inferred output variances, 
\begin{eqnarray}
S_{\text{inf}}^{\text{out}}(\hat{X}_{1}) & = & S_{X_{1}}^{\text{out}}-\frac{\left[V(\hat{X}_{1},\hat{X}_{2})\right]^{2}}{S_{X_{2}}^{\text{out}}},\nonumber \\
S_{\text{inf}}^{\text{out}}(\hat{Y}_{1}) & = & S_{Y_{1}}^{\text{out}}-\frac{\left[V(\hat{Y}_{1},\hat{Y}_{2})\right]^{2}}{S_{Y_{2}}^{\text{out}}}.
\label{eq:EPROPA}
\end{eqnarray}
The inferred variances for the $j=2$ quadratures are simply found
by swapping the indices $1$ and $2$. As the $\hat{X}_{j}$ and $\hat{Y}_{j}$
operators do not commute, the products of the variances obey a Heisenberg
uncertainty relation, with $S_{X_{j}}^{\text{out}}S_{Y_{j}}^{\text{out}}\geq1$.
Hence we find a demonstration of the EPR paradox whenever 
\begin{equation}
S_{\text{inf}}^{\text{out}}(\hat{X}_{j})S_{\text{inf}}^{\text{out}}(\hat{Y}_{j})\leq1.
\label{eq:demonstration}
\end{equation}
With the expressions for the variances given in Eq.~\ref{eq:coupledVXVY}
and the covariances of Eq.~\ref{eq:VXX}, we have all we need to
calculate the EPR correlations. Once again, however, the full expressions
are somewhat unwieldy, so we will present the results graphically.

\begin{figure}
\begin{center}\includegraphics[%
  width=0.90\columnwidth]{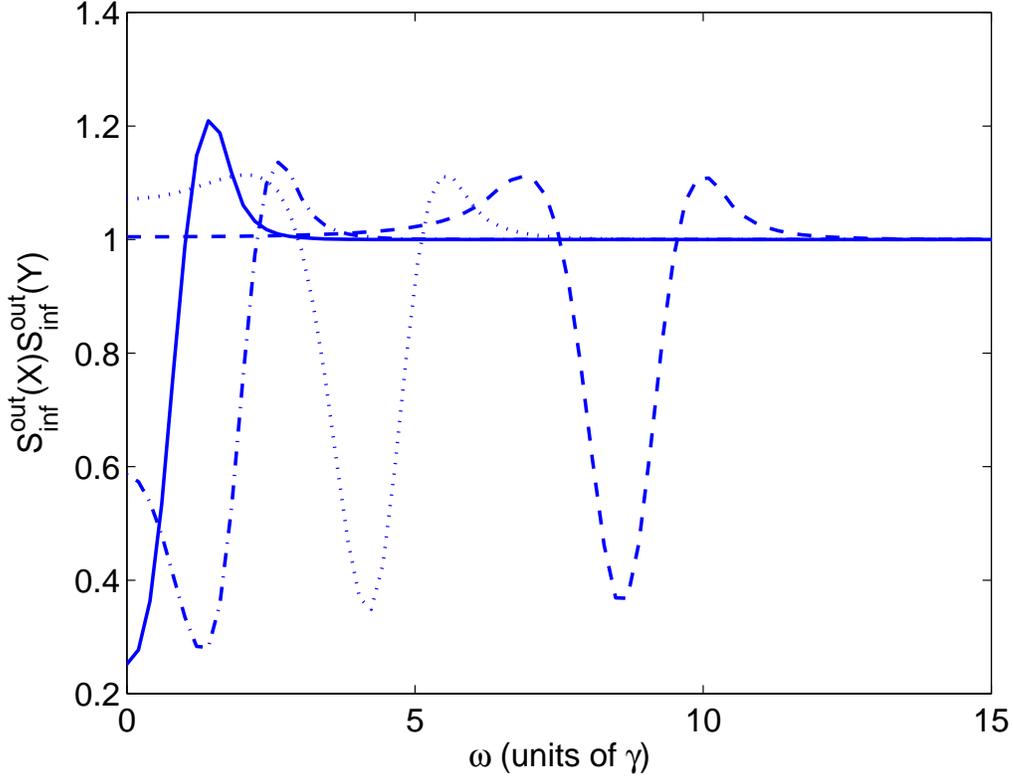}
\end{center}

\caption{Demonstration of the EPR correlation for $J_{b}=\gamma=1$ and $J_{a}=1$
(solid line, $\theta=67^{o}$), $2$ (dash-dotted line, $\theta=67^{o}$),
$5$ (dotted line, $\theta=113^{o}$) and $10$ (dashed line, $\theta=113^{o}$).
The pump amplitude is $\epsilon=0.5\epsilon_{c}$. }

\label{fig:epr1}
\end{figure}

In Fig.~\ref{fig:epr1} we present the results for optimised quadrature
phase angles while $J_{b}$ is held constant at a value of $\gamma$
while $J_{a}$ is increased. Note that again the angle $\theta$ refers
to the $\hat{X}^{\theta}$ quadratures, while the conjugate quadratures
are at an angle of $\theta+\pi/2$. Changing $J_{b}$ serves to change
the angle of the maximum violation, without changing the degree of
violation, while changing $J_{a}$ changes both the degree and the
frequency of the maximum violation. As expected, these results are
the same for both outputs of the device.

\section{Detuning the cavity}

\label{sec:detune}

Often in optical systems the best performance is found when the cavity
is resonant for the different modes involved in the interactions.
In the present case we find that detuning the cavity by the appropriate
amount from the two frequencies allows for some simplification of
the theoretical analysis and can actually improve some quantum correlations.
With detunings included, the steady state below threshold solutions
for the high frequency mode are found as 
\begin{equation}
\beta_{1}^{ss}=\beta_{2}^{ss}=\beta_{ss}=\frac{\epsilon}{\left[\gamma_{b}-i(J_{b}-\Delta_{b})\right]},
\label{eq:betadelta}
\end{equation}
so that, setting $\Delta_{b}=J_{b}$, we return to the well-known
real solutions for a single OPO. If we then set $\Delta_{a}=J_{a}$,
define the new variables $A_{p}=\alpha_{1}+\alpha_{2}$ and $A_{m}=\alpha_{1}-\alpha_{2}$,
and eliminate the time dependence of $\beta_{a,b}$, we can write
positive-P stochastic equations as 
\begin{eqnarray}
\frac{dA_{p}}{dt} & = & -\gamma_{a}A_{p}+\kappa\beta_{ss}A_{p}^{+}+\sqrt{\kappa\beta_{ss}}\left(\eta_{1}+\eta_{3}\right),\nonumber \\
\frac{dA_{p}^{+}}{dt} & = & -\gamma_{a}A_{p}^{+}+\kappa\beta_{ss}A_{p}+\sqrt{\kappa\beta_{ss}}\left(\eta_{2}+\eta_{4}\right),\nonumber \\
\frac{dA_{m}}{dt} & = & -\left[\gamma_{a}+2iJ_{a}\right]A_{m}+\kappa\beta_{ss}A_{m}^{+}+\sqrt{\kappa\beta_{ss}}\left(\eta_{1}-\eta_{3}\right),\nonumber \\
\frac{dA_{m}^{+}}{dt} & = & -\left[\gamma_{a}-2iJ_{a}\right]A_{m}^{+}+\kappa\beta_{ss}A_{m}+\sqrt{\kappa\beta_{ss}}\left(\eta_{2}-\eta_{4}\right).
\label{eq:PPcombine}
\end{eqnarray}
In the above, the noise terms are the same as those of Eq.~\ref{eq:PPSDE}.
We note here that, although it is the detuning in the low frequency
mode that allows us to write the equations for $A_{p}$ and $A_{p}^{+}$
in a particularly simple form, $\Delta_{b}$ also plays a role in
that it allows us to treat $\beta_{ss}$ as real, which will make
the interesting quantum correlations in and between the $X$ and $Y$
quadratures, so that we do not have to examine all possible local
oscillator angles to find the best performance.

Following the same linearisation procedure as in section~\ref{sec:linearise},
we find the corresponding drift and noise matrices, \begin{equation}
A_{pm}=\left[
\begin{array}{cccc}
\gamma_{a} & -\kappa\beta_{ss} & 0 & 0\\
-\kappa\beta_{ss} & \gamma_{a} & 0 & 0\\
0 & 0 & \gamma_{a}+2iJ_{a} & -\kappa\beta_{ss}\\
0 & 0 & -\kappa\beta_{ss} & \gamma_{a}-2iJ_{a}\end{array}\right],
\label{eq:Aplusminus}
\end{equation}
and \begin{equation}
B_{pm}=\left[
\begin{array}{cccc}
\sqrt{\kappa\beta_{ss}} & 0 & \sqrt{\kappa\beta_{ss}} & 0\\
0 & \sqrt{\kappa\beta_{ss}} & 0 & \sqrt{\kappa\beta_{ss}}\\
\sqrt{\kappa\beta_{ss}} & 0 & -\sqrt{\kappa\beta_{ss}} & 0\\
0 & \sqrt{\kappa\beta_{ss}} & 0 & -\sqrt{\kappa\beta_{ss}}\end{array}\right].
\label{eq:Bplusminus}
\end{equation}
In terms of the quadratures used in section~\ref{sec:correlations},
we now define 
\begin{eqnarray}
X_{p} & = & A_{p}+A_{p}^{+}=X_{1}+X_{2},\nonumber \\
X_{m} & = & A_{m}+A_{m}^{+}=X_{1}-X_{2},\nonumber \\
Y_{p} & = & -i\left(A_{p}-A_{p}^{+}\right)=Y_{1}+Y_{2},\nonumber \\
Y_{m} & = & -i\left(A_{m}-A_{m}^{+}\right)=Y_{1}-Y_{2},
\label{eq:quadcombine}
\end{eqnarray}
and give expressions for the output spectral variances of these new
quadratures. For the $X_{p}$ and $Y_{p}$ quadratures these are particularly
simple, 
\begin{eqnarray}
S_{X_{p}}^{\text{out}}(\omega) & = & 2+\frac{8\gamma_{a}\gamma_{b}\kappa\epsilon}{\left(\gamma_{a}\gamma_{b}-\kappa\epsilon\right)^{2}+\gamma_{b}^{2}\omega^{2}},\nonumber \\
S_{Y_{p}}^{\text{out}}(\omega) & = & 2-\frac{8\gamma_{a}\gamma_{b}\kappa\epsilon}{\left(\gamma_{a}\gamma_{b}+\kappa\epsilon\right)^{2}+\gamma_{b}^{2}\omega^{2}},
\label{eq:spekplus}
\end{eqnarray}
and are readily seen to be the sum of the variances for two uncoupled
OPOs, as given in Eq.~\ref{eq:noncoupledVXVY}. As in that case,
the zero-frequency variance in $Y_{p}$ is predicted to vanish at
the critical pump value of $\epsilon_{c}=\gamma_{a}\gamma_{b}/\kappa$,
although, as should be well known, a linearised analysis is not valid
in that region. However, the degree of squeezing is more than was
found to be available in the doubly resonant case considered above.
The other two variances do not uncouple and have more complicated
expressions, 
\begin{eqnarray}
S_{X_{m}}^{\text{out}}(\omega) & = & 2+\frac{8\gamma_{a}\gamma_{b}\kappa\epsilon\left[(\gamma_{a}\gamma_{b}+\kappa\epsilon)^{2}-\gamma_{b}^{2}(4J_{a}^{2}-
\omega^{2})\right]}{\left[\gamma_{b}^{2}(\gamma_{a}^{2}+4J_{a}^{2}-\omega^{2})-\kappa^{2}\epsilon^{2}\right]^{2}+4\gamma_{a}^{2}\gamma_{b}^{4}\omega^{2}},\nonumber \\
S_{Y_{m}}^{\text{out}}(\omega) & = & 2-\frac{8\gamma_{a}\gamma_{b}\kappa\epsilon\left[(\gamma_{a}\gamma_{b}-\kappa\epsilon)^{2}-\gamma_{b}^{2}(4J_{a}^{2}
-\omega^{2})\right]}{\left[\gamma_{b}^{2}(\gamma_{a}^{2}+4J_{a}^{2}-\omega^{2})-\kappa^{2}\epsilon^{2}\right]^{2}+4\gamma_{a}^{2}\gamma_{b}^{4}\omega^{2}}.\nonumber \\
\label{eq:spekminus}
\end{eqnarray}

\begin{figure}
\begin{center}\includegraphics[%
  width=0.90\columnwidth]{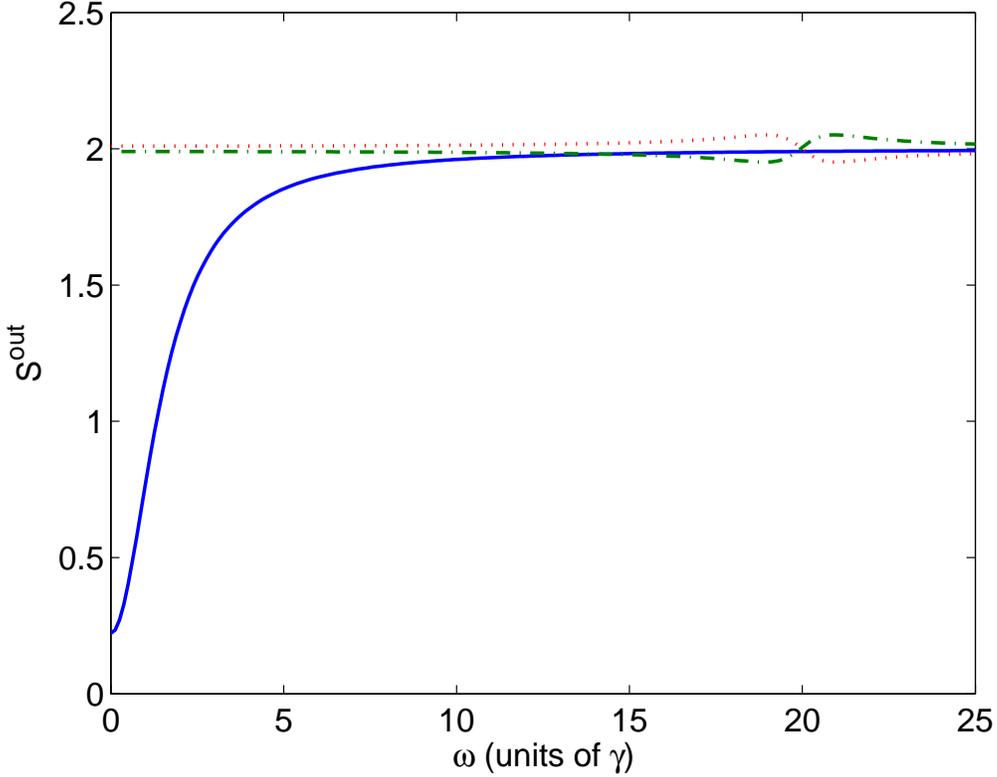}\end{center}

\caption{Output spectral variances of the combined modes for $J_{b}=\Delta_{b}=\gamma_{a}=\gamma_{b}=\gamma=1$
and $J_{a}=\Delta_{a}=10$. The solid line is $S_{Y_{p}}^{\text{out}}$,
the dash-dotted line is $S_{X_{m}}^{\text{out}}$, and the dotted
line is $S_{Y_{m}}^{\text{out}}$. A value of less than $2$ represents
squeezing. The pump amplitude is $\epsilon=0.5\epsilon_{c}$.}

\label{fig:combvar}
\end{figure}

Graphical results for the combined quadratures which exhibit squeezing
are shown in Fig.~\ref{fig:combvar}, from which it is obvious that
by far the best squeezing quadrature is $Y_{p}$, which, for these
parameters, shows almost $90\%$ squeezing at zero-frequency. The
quadratures $X_{m}$ and $Y_{m}$ show only a very small degree of
squeezing far from zero frequency. What this result shows, along with
the results for the resonant cavity, is that the low frequency modes
want to oscillate at two distinct frequencies, as is normal for coupled
systems. The detuning chosen, $\Delta_{a}=J_{a}$, moves the sum mode
frequency closer to resonance while the other frequency is further
detuned. Along with the choice of $\Delta_{b}$ so as to make the
intracavity high frequency amplitude real, this results in maximised
single-mode noise supression and entanglement centred on zero frequency.

\begin{figure}
\begin{center}\includegraphics[%
  width=0.90\columnwidth]{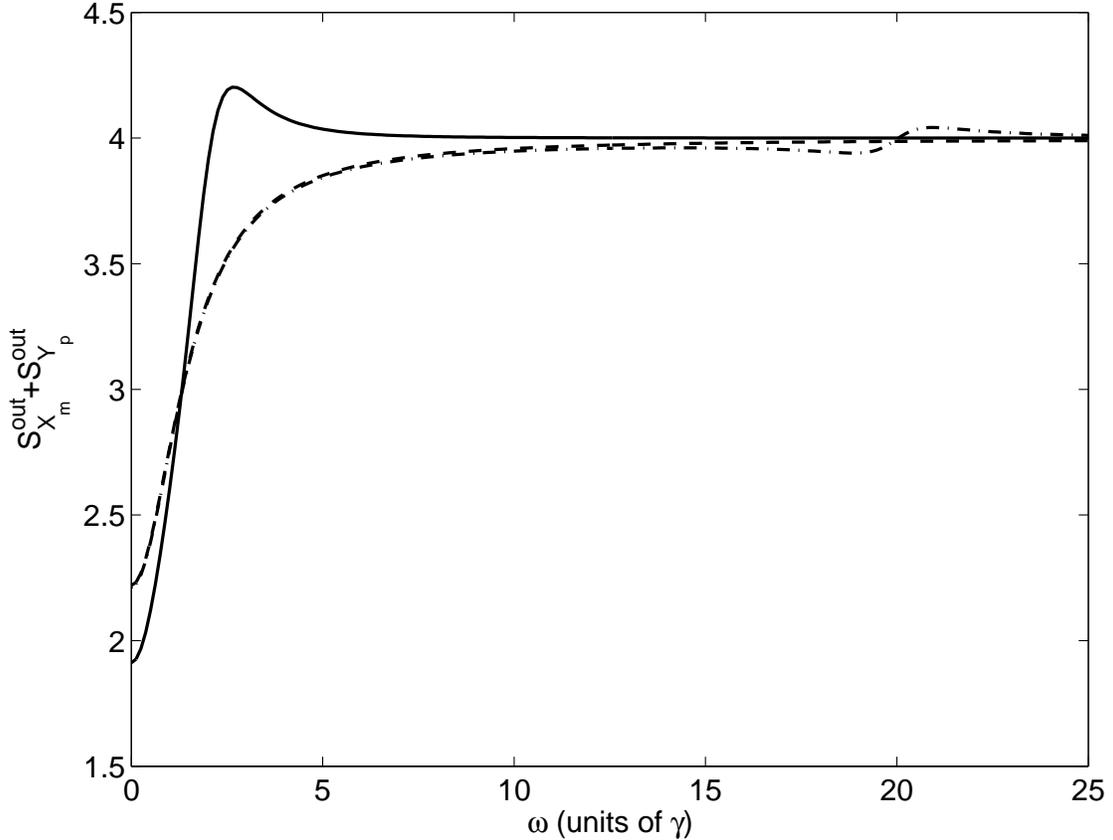}\end{center}

\caption{The sum of the output spectral variances, $S_{X_{m}}^{\text{out}}+S_{Y_{p}}^{\text{out}}$,
for $J_{b}=\Delta_{b}=\gamma_{a}=\gamma_{b}=\gamma=1$ and $J_{a}=\Delta_{a}=1$
(solid line), $10$ (dash-dotted line) and $20$ (dashed line). A
value of less than $4$ represents entanglement. The pump amplitude
is $\epsilon=0.5\epsilon_{c}$.}

\label{fig:comentangle}
\end{figure}

\begin{figure}
\begin{center}\includegraphics[%
  width=0.90\columnwidth]{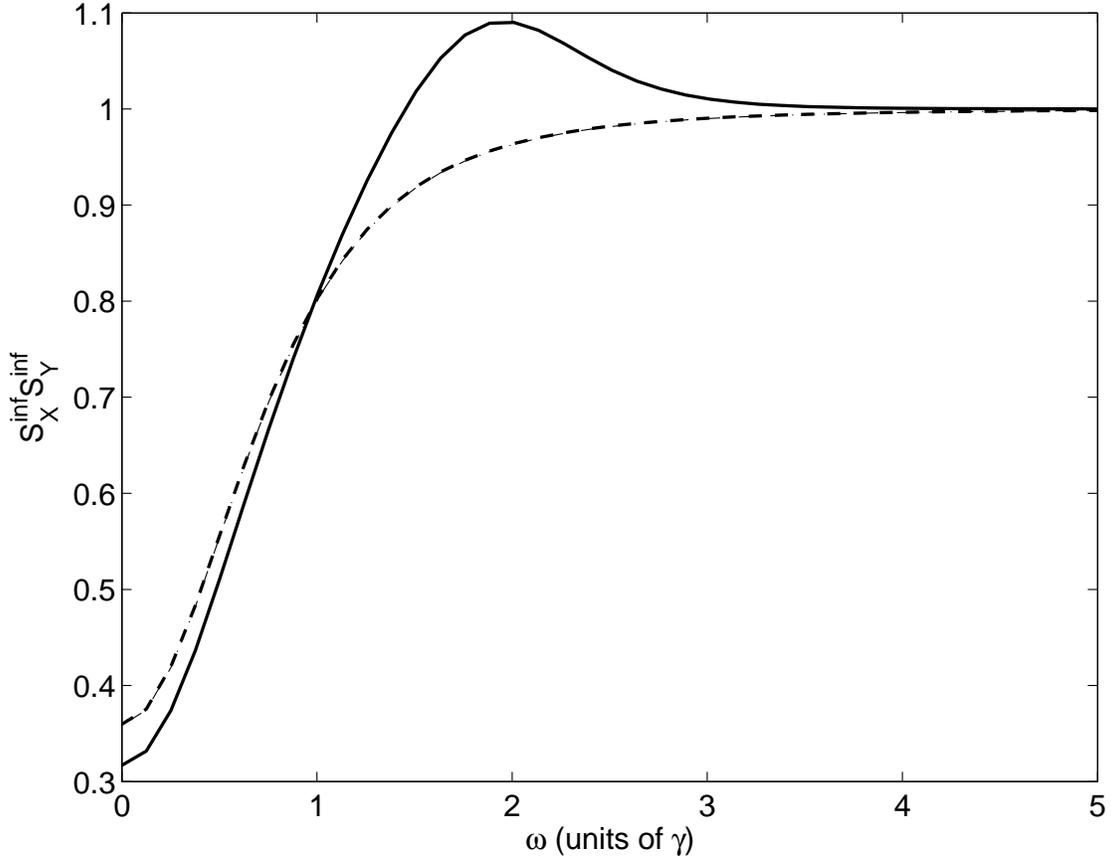}\end{center}

\caption{The product of the inferred output spectral variances, $S_{X}^{inf}S_{Y}^{inf}$,
for $J_{b}=\Delta_{b}=\gamma_{a}=\gamma_{b}=\gamma=1$ and $J_{a}=\Delta_{a}=1$
(solid line) and $10$ (dash-dotted line). On this scale, the result
for $J_{a}=20$ is indistinguishable from that for $J_{a}=10$. A
value of less than $1$ represents a demonstration of the EPR paradox.
The pump amplitude is $\epsilon=0.5\epsilon_{c}$.}

\label{fig:detepr}
\end{figure}

Using these results, we can now investigate the degree of entanglement,
as done above for the resonant case. As shown in Fig.~\ref{fig:comentangle},
we find that the quadratures $Y_{p}$ and $X_{m}$ are entangled,
exactly as in the single OPO case. As with the squeezing, the detunings
have moved the maximum of entanglement to zero frequency. A sign of
the out of resonance mode attempting to resonate is seen in the small
degree of entanglement apparent around $\omega\approx20\gamma$. We
also see that the degree of entanglement is less than in the case
with zero detuning, shown previously in Fig.~\ref{fig:entangleJa},
although it must be remembered that the absolute pump powers are not
the same, merely the ratios $\epsilon/\epsilon_{c}$. Finding analytical
expressions for EPR correlations is not possible using this coupled-mode
approach, as, although we can calculate the necessary covariances,
for example, $V(\hat{X}_{1},\hat{X}_{2})=[V(X_{p})-V(X_{m}))]/4$,
it is not obvious how to separate out the single-mode variances. However,
these can still be calculated numerically using the full single-mode
equations with the appropriate detunings. That the system clearly
demonstrates the EPR paradox is shown in Fig.~\ref{fig:detepr},
although again the maximum inferred violation is less than in the
resonant case.

We note here that all the quantities shown for the detuned system
are actually calculated at a lower absolute pump power than in the
resonant case. For positive detunings, the critical pump amplitude
is found as 
\begin{equation}
\epsilon_{c}=\sqrt{[\gamma_{a}^{2}+(J_{a}-\Delta_{a})^{2}][\gamma_{b}^{2}+(J_{b}-\Delta_{b})^{2}]}/\kappa\,\,,
\end{equation}
so that our choice of detunings means this is no longer a function
of the coupling strengths. Therefore a careful choice of detunings
has two main advantages in that it fixes the quadratures for which
the maxima of quantum features are found, and means that the pumping
necessary to a good performance does not vary with the coupling strengths. 

The choice of detunings shown has the possible disadvantage that,
as the effective coupling is now only in the $A_{m}$ mode, which
is moved away from resonance, the quantum correlations which depend
on both the modes are slightly degraded. This is readily seen from
the figures because those correlations which include $X_{m}$ and
$Y_{m}$ change more with $J_{a}$ than do the others.

\section{Conclusion}

This system exhibits a wide range of behaviour and is potentially
an easily tunable source of single-mode squeezing, entangled states
and states which exhibit the EPR paradox. The spatial separation of
the output modes means that they do not have to be separated by optical
devices before measurements can be made, along with the unavoidable
losses which would result from this procedure. The entangled beams
produced can be degenerate in both frequency and polarisation, unlike
those of the nondegenerate OPO, and would exit the cavity at spatially
separated locations. This may be a real operational advantage over
the nondegenerate OPO, which is also known to produce nonclassical
states. The tunability that exists because of the number of different
parameters which can be experimentally accessed, such as the coupling
strength, the pump intensity and the detunings, may make it interesting
for a range of potential applications which would require the availability
of states of the electromagnetic field with varying degrees of nonclassicality.
Since this type of system is compatible with integrated optics techniques,
it may provide a more robust source of entanglement than interferometers
that use discrete optical components.

\begin{acknowledgments}
This research was supported by the Australian Research Council.
\end{acknowledgments}

\end{document}